\documentclass[emulateapj,natbib]{emulateapj}

\begin{document}

\title{Direct Imaging and Spectroscopy of a Young Extrasolar Kuiper Belt in the Nearest OB Association}
\author{
Thayne Currie\altaffilmark{1}, 
Carey M. Lisse\altaffilmark{2},
Marc Kuchner\altaffilmark{3},
Nikku Madhusudhan\altaffilmark{4},
Scott J. Kenyon\altaffilmark{5},
Christian Thalmann\altaffilmark{6},
Joseph Carson\altaffilmark{7},
John Debes\altaffilmark{8}
}
\altaffiltext{1}{National Astronomical Observatory of Japan}
\altaffiltext{2}{Applied Physics Laboratory, The Johns Hopkins University}
\altaffiltext{3}{NASA-Goddard Space Flight Center}
\altaffiltext{4}{Institute for Astronomy, University of Cambridge}
\altaffiltext{5}{Harvard-Smithsonian Center for Astrophysics}
\altaffiltext{6}{ETH-Zurich}
\altaffiltext{7}{Department of Physics and Astronomy, The College of Charleston}
\altaffiltext{8}{Space Telescope Science Institute}

\begin{abstract}
We describe the discovery of a bright, young Kuiper belt-like debris disk around HD 115600, a $\sim$ 1.4--1.5 M$_\mathrm{\odot}$, $\sim$ 15 Myr old member of the Sco-Cen OB Association.  Our H-band coronagraphy/integral field spectroscopy from the \textit{Gemini Planet Imager} shows the ring has a (luminosity scaled) semi major axis of ($\sim$ 22 AU) $\sim$ 48 AU, similar to the  current Kuiper belt. 
The disk appears to have neutral scattering dust, is eccentric (e $\sim$ 0.1--0.2), and could be sculpted by analogues to the outer solar system planets. 
  Spectroscopy of the disk ansae reveal a slightly blue to gray disk color, consistent with major Kuiper belt chemical constituents, where water-ice is a very plausible dominant constituent. 
    Besides being the first object discovered with the next generation of extreme adaptive optics systems (i.e. SCExAO, GPI, SPHERE), HD 115600's debris ring and planetary system provides a key reference point for the early evolution of the solar system, the structure and composition of the Kuiper belt, and the interaction between debris disks and planets.
\end{abstract}
\section{Introduction}
Situated just beyond Neptune's orbit (30--50 AU), the Kuiper belt is home to thousands of remnants of the earliest stages of icy planet formation, is the location of numerous dwarf planets such as Pluto, and provides keys to understanding the early, unevolved solar system  \citep{Jewitt1992}.   Cold debris rings around nearby, young stars offer a critical reference for the Kuiper belt's evolution and composition \citep{KenyonBromley2008,Wyatt2008}.  However, the few such rings imaged in scattered light surround stars at stellocentric distances farther than the Kuiper belt \citep[e.g.][]{Kalas2005,Schneider2009,Krist2012}, are not located in massive OB association resembling the solar system's likely birth environment \citep{Adams2010}, and lack spatially-resolved scattered-light spectroscopy to explore composition.    

HD 115600 \citep[$d$ = 110.5 $pc$;][]{vanLeewen2007} is a 1.4 M$_\mathrm{\odot}$, F2V/F3V 15 $Myr$-old member of the Lower Centaurus Crux (LCC) region of the Sco-Cen OB Association \citep{Pecaut2012} \footnote{Masses and ages were derived using the \citet{Baraffe2015} isochrones.   Slightly higher masses with the same 15 Myr age are also consistent \citep[e.g. 1.5 M$_\mathrm{\odot}$][]{Pecaut2012}.}.  The star has a large infrared excess with a dust mass and fractional luminosity (0.05 $M_\mathrm{moon}$; $L_\mathrm{IR}$/$L_\mathrm{\star}$ $\sim$ 1.7 $\times$10$^{-3}$) comparable to the well-known luminous, resolved ring around HR 4796A \citep{Chen2011,Chen2015,Jura1998}.  Thus, HD 115600 is a promising target for high-contrast imaging.

We report the discovery of a bright debris ring around HD 115600 using integral field spectroscopy from the \textit{Gemini Planet Imager} \citep{Macintosh2014}.  The ring is plausibly sculpted by an unseen solar system-like planet, surrounds the star at a Kuiper belt-like distance, and has a spectrum consistent with major Kuiper belt constituents.  The HD 115600 planetary system provides a valuable reference point for the early evolution and composition of the solar system, the structure of the early Kuiper belt, and the interaction between debris disks and planets.  
\\
\section{Observations and Data Reduction}

 We obtained integral field spectroscopy of HD 115600 on 23 March 2014 with the \textit{Gemini Planet Imager} 
\citep{Macintosh2014} in $H$ band ($\lambda_{o}$ = 1.6 $\mu$m, $R$ = 44--49; 0\farcs{}014 pixel$^{-1}$) behind the apodized pupil lyot coronagraph 
($r_{mask}$ $\sim$ 0\farcs{}12) and in angular differential imaging mode \citep[ADI;][]{Marois2006}.  
Our observations consisted of 58 co-added 49.5 second frames, yielding a total integration time of 
$\approx$ 48 minutes and a field rotation of  29$^\circ$ (3 $\lambda$/D at 0\farcs{}25).   
\textit{Xe} and \textit{Ar} lamp observations provided wavelength calibration.

We carried out basic data reduction using the 
\textit{Gemini Data Reduction Pipeline, version 1.2.1} \citep{Perrin2014}.  We modified the software, adding image registration steps as in \citet{Currie2011a}.
Removing the middle 20 data cubes, which exhibited a pupil stop misalignment, and those with poorer AO correction reduces our total integration time to 28 minutes.

 For point-spread function (PSF) subtraction, we used A-LOCI \citep{Currie2012,Currie2014c} tuned to ``conservative" settings \citep[][]{Thalmann2011} better suited for recovering extended emission, imposing a rotation gap criterion of $\delta$ = 1.5, an optimization area of $N_\mathrm{A}$ = 1000, 
and a \textit{singular value decomposition} (SVD) cutoff of \textit{SVD$_\mathrm{lim}$} = 10$^{-5}$ 
\citep[see ][]{Lafreniere2007a,Currie2014b}.    
We construct a final data cube from a median-combination of PSF-subtracted cubes and wavelength-collapsed using an outlier-resistant mean to yield a band-averaged image.   
Due to the small number of data cubes, we did not employ speckle filtering \citep{Currie2012,Currie2014b} to truncate the set of reference PSFs.  
For initial flux calibration, we measured the average, background-subtracted signal of the four satellite spots (transmission =  2.035 $\times$ 10$^{-4}$) in an 8-pixel diameter aperture and divided the data cube by a normalized F2V spectrum \citep{Pickles1998}.  

\section{Detection of the HD 115600 Debris Disk}

\begin{figure}
\centering
\includegraphics[scale=0.4,trim=40mm 15mm 40mm 15mm,clip]{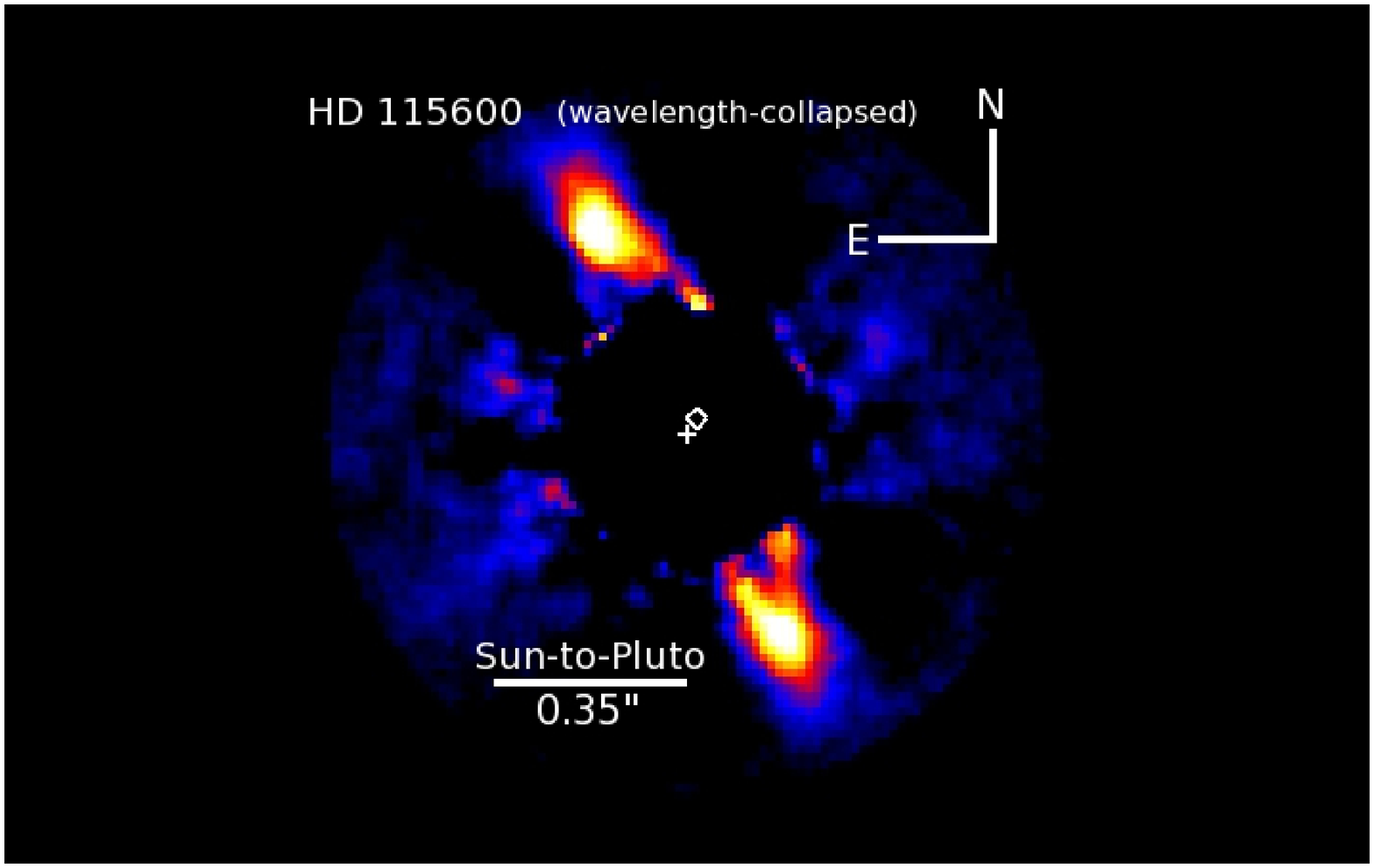}
\includegraphics[scale=0.4,trim=40mm 15mm 40mm 15mm,clip]{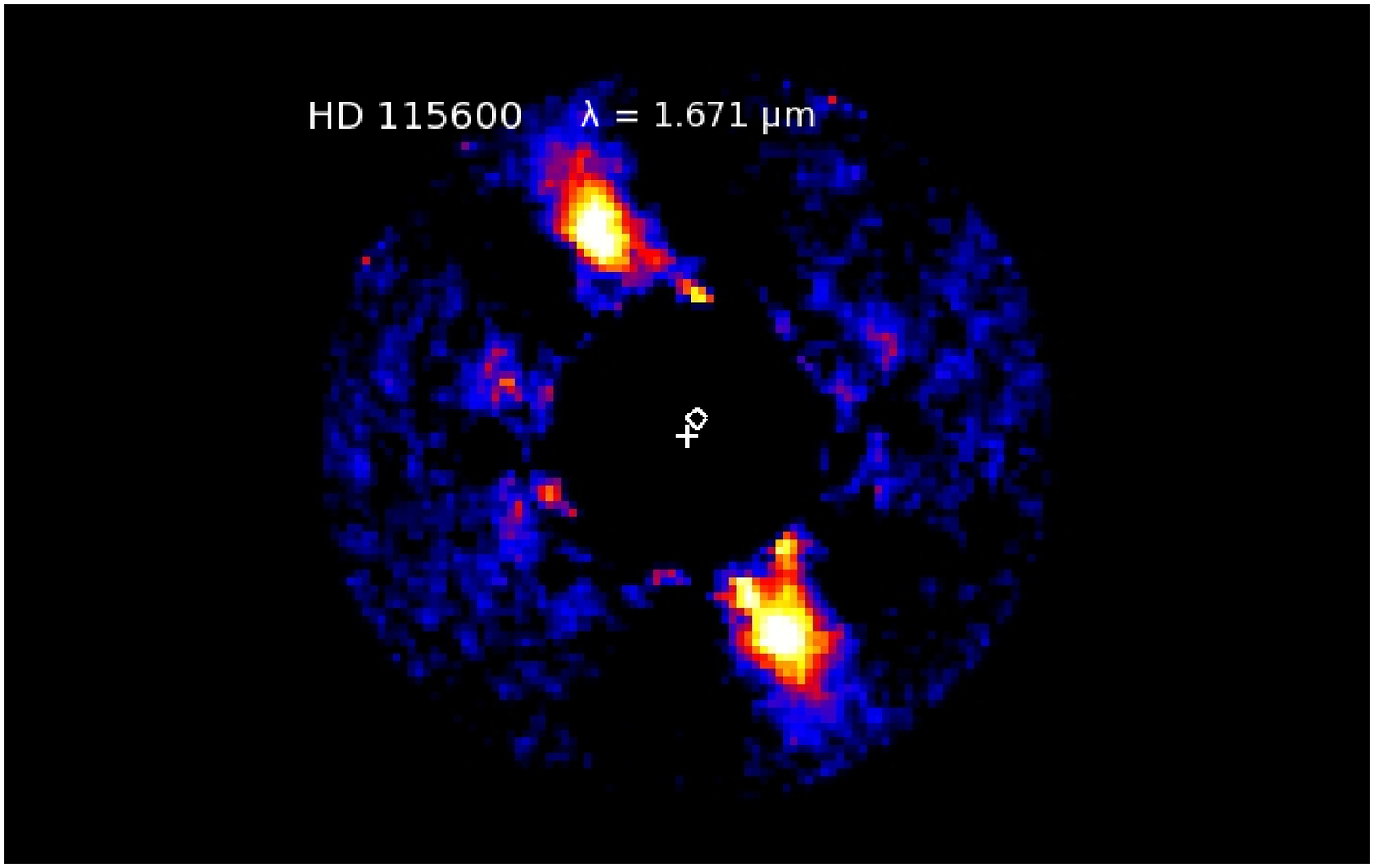}\\
\includegraphics[scale=0.4,trim=40mm 15mm 40mm 15mm,clip]{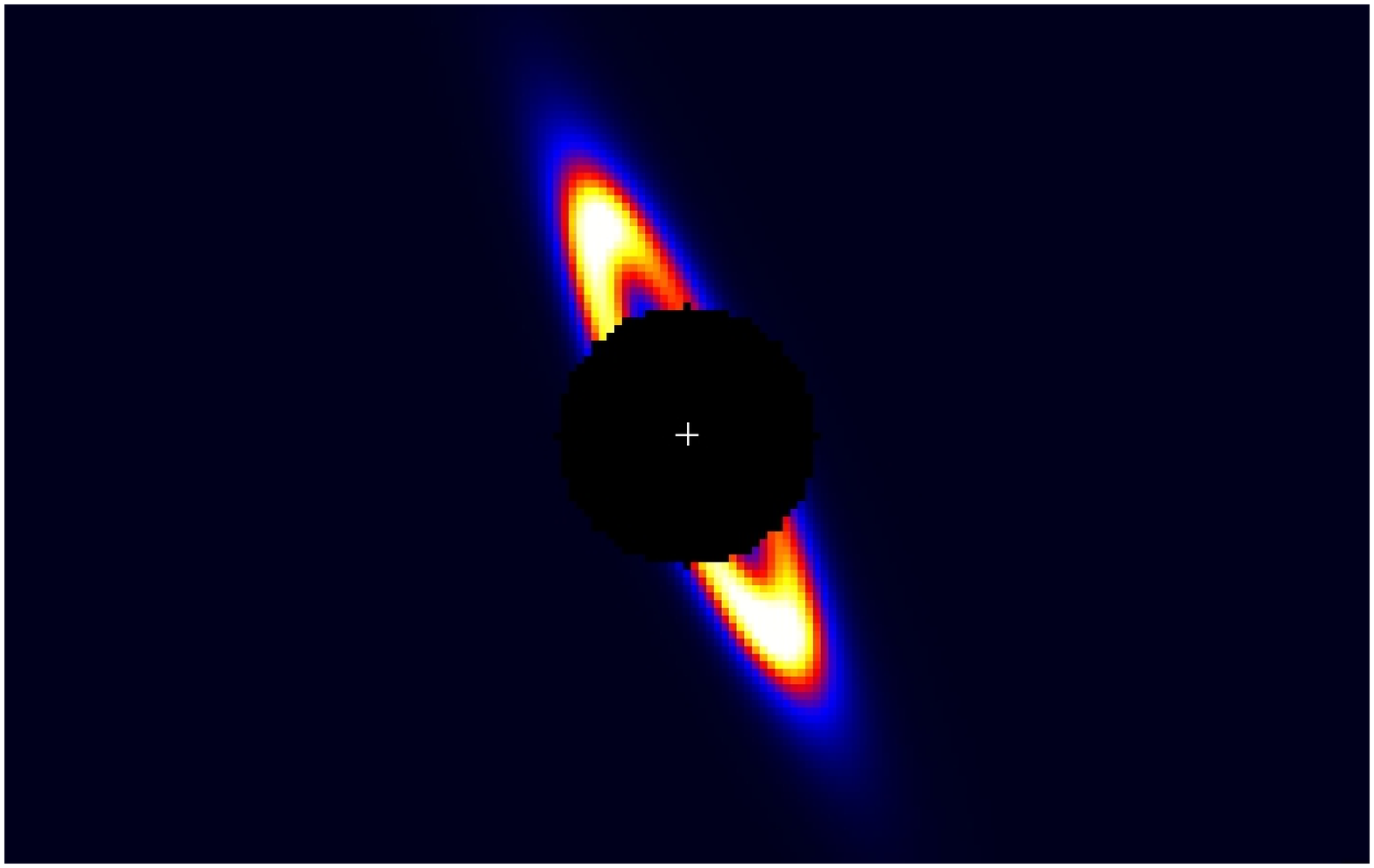}
\includegraphics[scale=0.4,trim=40mm 15mm 40mm 15mm,clip]{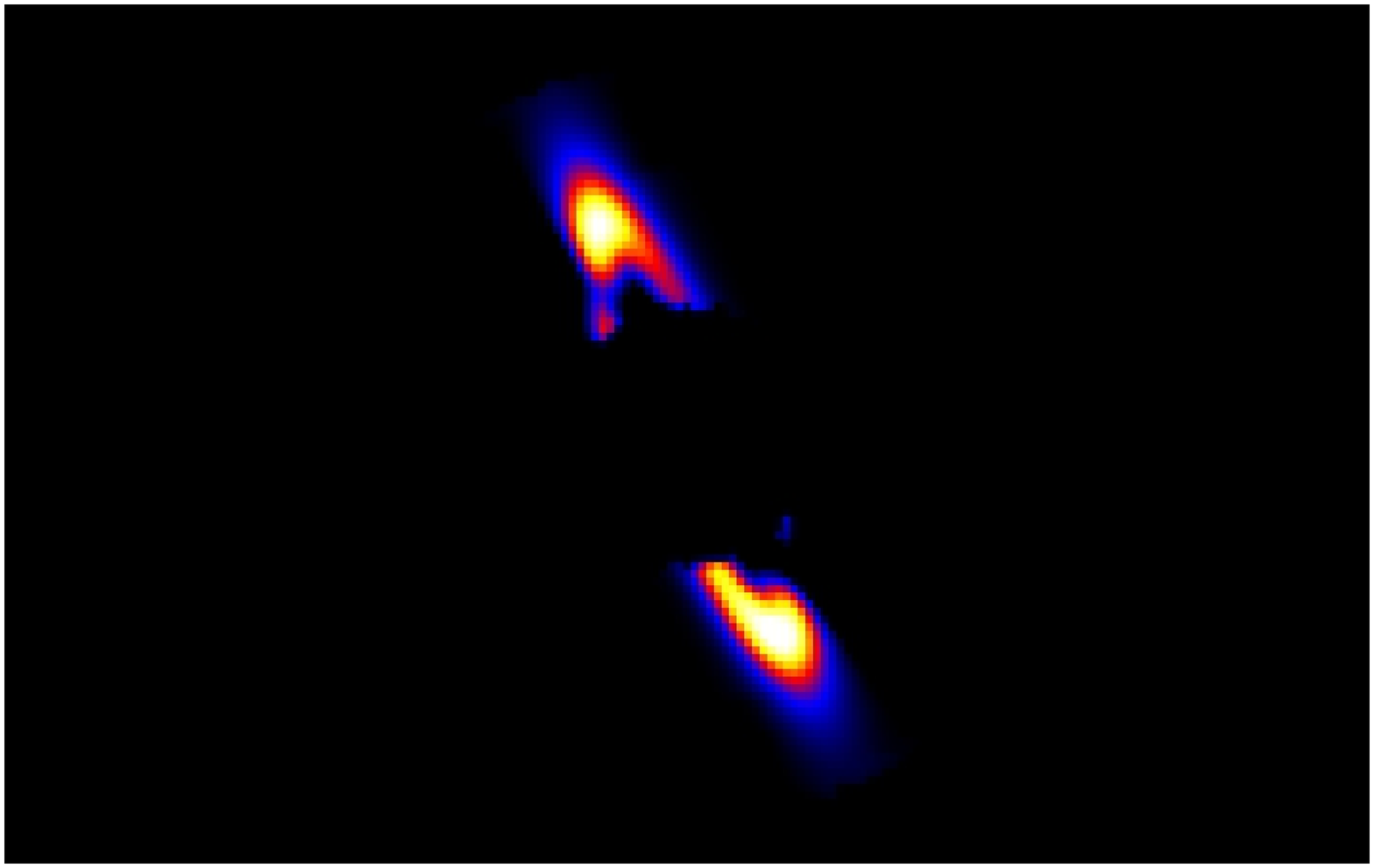}
\caption{(top-left) Reduced, PSF-subtracted, and wavelength-collapsed image of HD 115600 showing a bright, nearly edge-on debris ring located just beyond a Pluto-like distance to the star.  The disk center (diamond), determined from ellipse fitting, appears offset from the star's position (cross). (top-right) A representative wavelength slice (1.671 $\mu$m) of the final datacube.   
(bottom-left) The best-fit model disk produced by GRaTeR:  a$_{o}$ = 48 AU, $PA$ = 24$^\circ$, $i$ = 79.5$^\circ$, $\theta_{o}$ = 90$^\circ$, $ksi_{o}$ = 3 AU, $g$ = 0, $e$ = 0.2, $\alpha_{in}$ = $7.5$, and $\alpha_{out}$ = $-5$.  (bottom-right) The same model, convolved with the appropriate PSF in each channel, PSF subtracted, and wavelength-collapsed.  The model reproduces the suppressed signal in the northeast and southwest portions of the disk, although there are some small discrepancies (see main text).
}
\label{images}
\end{figure}

Figure \ref{images} (top panels) displays our wavelength-collapsed disk image and an individual channel, clearly revealing a bright, thick debris ring visible from $r$ $\sim$  0\farcs{}25 to $r$ $\sim$ 0\farcs{}55 and viewed close to edge-on.   The disk ansae along the major axis extend from $r$ $\sim$ 0\farcs{}34 to 0\farcs{}5, or 37.5 AU to 55 AU, comparable to the current Kuiper belt.
Both sides of the disk are visible, ruling out strong forward-scattering anisotropy.   The disk is offset from the star position and  
visually resembles a puffier version of the HR 4796A disk \citep{Schneider2009}. 

 To estimate the disk's SNR in the collapsed image and in each spectral channel, we consider two methods.  First, for the collapsed image we construct a SNR map as in \citet{Currie2011a},  replacing each pixel with the sum of values enclosed by a 3-pixel aperture ($\sim$ 1 FWHM) and computing the robust standard deviation at each angular separation.   For a given region of the disk, this procedure considers other disk regions (e.g. those on the opposite side) and negative self-subtraction residuals as ``noise", which leads to underestimating the significance of the disk signal compared to real residual speckle noise.  Nevertheless, the disk ansae in the collapsed image are still detected at SNR $\sim$ 8; the disk's signal exceeds 2-$\sigma$ exterior to $r$ $\sim$ 0\farcs{}25.    
 
  Second, we compute the ansae SNR in each spectral channel, defining the noise likewise as the robust standard deviation of the summed image in a FWHM-wide arc at the same angular separation as the disk but at different position angles, at a separation (10--50 pixels away) chosen to avoid the disk signal and negative self-subtraction footprints.   The SNR in each spectral channel in the disk ansae ranges between 5 and 8; the SNR of the collapsed image exceeds 10 in the ansae.  

\section{Analysis}
To derive the disk geometry and extract spectra, we follow a three-step approach.  First, we derive the debris ring's basic geometry from the wavelength-collapsed data cube using ellipse fitting.  Second, we use forward-modeling to fine-tune these properties and calculate second-order properties of the disk (e.g. scattering function).   Third, we extract the disk's spectrum in the bright ansa regions, using the best-fit model to the final data cube to calibrate the disk throughput in each spectral channel.
\subsection{Geometry}
To determine the basic geometry of the disk -- inclination, position angle, semi-major/minor axes, and disk center -- we first use the 
using the IDL \textrm{mpfitellipse} package to determine an approximate trace of the disk, where the pixels are weighted by their SNR.   Second, we constructed a grid of ellipse parameters around the best-fit set determined by \textrm{mpfitellipse}, and calculated a more robust best-fit value using the ``maximum merit" procedure, identifying the ellipse parameters that maximized the disk signal along the trace of the ellipse \citep{Thalmann2011}.  We repeat this step using different ranges in radii and different cutoffs in SNR for the disk trace (e.g. SNR $>$ 1.5, 2; $r$ = 17.5--45 pixels; 20--40 pixels) to define best-estimated values and associated uncertainties.

We derive a best-fit position angle and inclination of PA = 24$^\circ$ $\pm$ 0.5$^\circ$ and  $i$ = 79.5$^\circ$ $\pm$ 0.5$^\circ$ (1-$\sigma$ errors).    The disk semimajor/minor axes are $r_{major, minor}$ = 0\farcs{}425 $\pm$ 0\farcs{}010 ($\sim$ 48 AU $\pm$ 1.1 AU), 0\farcs{}077 $\pm$ 0\farcs{}007 (8.5 $\pm$ 0.8 AU).  The semimajor axis and visible extent of the disk ansae are within the range of radii encompassing the current Kuiper belt; the luminosity-scaled ($r_{major}$/$\sqrt{L_{\star}}$) semi major axis (extent of the ansae) are $\sim$ 22 AU (17--25 AU), comparable to the predicated distances of major models of the early, pre-stirred Kuiper belt,: e.g. the Nice model  \citep[][]{Levison2008} or \citet{Nesvorny2015}.   The projected disk center is offset from the star $\Delta$x,$\Delta$y = 0\farcs{}018 $\pm$ 0\farcs{}008, 0\farcs{}029 $\pm$ 0\farcs{}014 (Fig. 1; diamonds) and the peak signal of the two ansae differ by $\sim$ 0.5 pixel in angular separation ($\sim$ 1 AU).

\subsection{Disk Forward Modeling}
To infer additional disk properties, we generate a grid of synthetic scattered light images using the GRaTeR code \citep{Augereau1999}, and forward-model these synthetic disks using our pipeline to compare the processed synthetic disk image with the real disk image in each spectral channel, extending to integral field spectroscopy methods that have been applied to broadband imaging data \citep{Esposito2014,Mazoyer2014}.
 We insert a disk model into a sequence of empty data cubes with position angles identical to those of our science sequence, and convolve the model in each spectral channel with the appropriately-sized PSF.     We performed PSF subtraction on the cubes containing the model disk using the same A-LOCI coefficients that were applied to the real data.  

Table \ref{diskmodels} describes the model parameter space we explore and our results.  To reduce the dimensionality of our forward-modeling, we adopt the inclination and position angle determined from our ellipse modeling (79.5$^\circ$ and 24$^\circ$, respectively).  
We defined the argument of pericenter to be [90, 270]$^\circ$, or 90$^\circ$ from the visible disk major axis:  departures from these values resulted in brightness asymmetries between the disk ansae inconsistent with the data.   
We varied other parameters, including the the ring center ($a_{o}$ = 47 and 48 AU), the Henyey-Greenstein scattering parameter $g$ (0--0.15), the disk scale height at the disk center ($ksi_{o}$ = 1--5 AU), the density power laws describing the decay of ring emission away from the ring center ($\alpha_{in}$ = 7.5, 10; $\alpha_{out}$ = $-$5, $-$7.5),  the disk eccentricity (0--0.3), and the offset along the major axis (0 and 1 AU).     While our parameter space search is not exhaustive, values outside these ranges (e.g. $g$ $>$ 0.15) yielded processed synthetic disk images strongly discrepant with the real data.
  
To match the observed brightness in each spectral channel, we fix the
parameters in the model and scale the flux and define the fit of the model to the data using the real and synthetic collapsed images convolved by the PSF as in our SNR calculations.   
Our ``acceptably fitting" models are those fulfilling $\chi^{2}$ $\le$ $\chi^{2}_{min}$ + $\sqrt{2\times N_{data}}$ \citep{Thalmann2013}.  We perform our optimization (and define $\chi^{2}$) by the visible trace of the disk at separations where it is consistently visible at SNR $>$ 2 along both sides ($\sim$ 0\farcs{}28 and 0\farcs{}52). 

The best-fit model is an offset, eccentric ring with neutral-scattering dust (Figure \ref{images} (bottom panels)).   
This model exhibits some discrepancies with the data, slightly over (under) predicting the signal on the southeastern (southwestern) side.  However, our best-fit model generally reproduces disk shape, providing a good first investigation.  More sophisticated models fit to higher SNR data and exploring more parameter space will better constrain disk properties.
Additionally, much of our parameter space remains degenerate, admitting widely varying values.   
However, we strongly prefer models with $g$ = 0: those with $g$ $>$ 0.05 predict disk ansae that are too faint compared to regions at small angular separations.  Models with $e$ $<$ 0.1 or $e$ = 0.3 fail to properly trace the observed disk.  A single model with $e$ = 0.1 fits our $\chi^{2}$ threshold, while all other acceptedly-fitting models have $e$=0.2, suggesting a disk eccentricity ($e$ = 0.1--0.2) comparable to that of the most eccentric known debris disks \citep[][]{Kalas2005,Krist2012}.   The deprojected, deconvolved, and azimuthally-averaged best-fit disk model has a normalized FWHM of 0.37, among the largest for scattered light-imaged debris rings \citep{Rodigas2015}.

\subsection{Dynamical Sculpting of the Disk: Limits on Planets}
Debris disk dynamical modeling sets stronger limits on unseen planets than from the GPI contrasts limits alone \citep[contrast $>$ 10$^{-5}$ at r $<$ 0\farcs{}4 (44 AU) or $M_{\mathrm{planet}}$ $>$ 7 $M_\mathrm{J}$; c.f.][]{Baraffe2003}.
The eccentric ring is plausibly due to dynamical perturbations from an unseen planet, which can open a gap in the disk and stir it via secular perturbations.   In both cases, the planet's eccentricity can be estimated assuming that the eccentricity of the ring equals the local value of the forced eccentricity \citep{Quillen2006}.  First, we place limits on the properties of the planet sculpting the disk from the \citet{Nesvold2015} gap opening model, where the planetesimals are initially dynamically hot ($0.0$ $<$ $e$ $<$ $0.2$) and the planet interacts with the disk via mean-motion resonances, triggering collisions and producing an inner hole.  Gap opening depends on the system age in units of the grain collisional time, which in turn depends on the orbital period and optical depth: $t_\mathrm{orbit}$/(4$\pi$$\tau$).  

Second, we explore the \citet{Mustill2009} planet stirring model: the planet stirs an initially cold disk from the inside-out via secular interactions, and the radius of the central hole extends to where planetesimal orbits begin to cross.  We compute stirring from first-order secular theory (valid for low planet eccentricities), corresponding here to larger planet semimajor axes.  Our two calculations assume 48 $AU$ as the ring center, the fractional luminosity as a proxy for optical depth, and an age of 15 Myr.   

\begin{figure}
\centering
\includegraphics[scale=0.5,clip,angle=0]{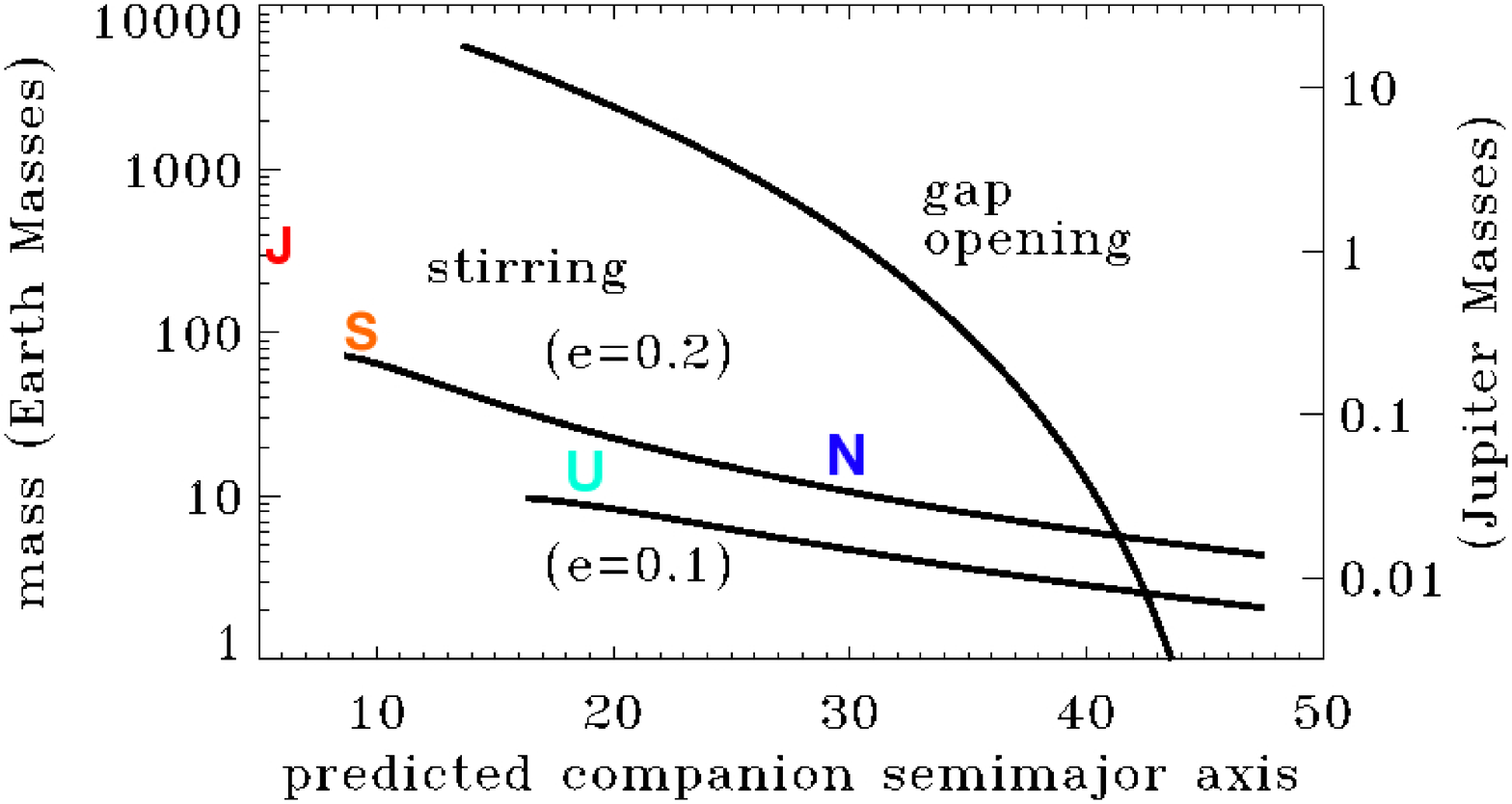}
\caption{Limits on the mass of a planet stirring ring planetesimals or opening a gap in the disk and clearing material out into a ring.   
Planets much like the outer solar system planets (overplotted) can stir a disk with an eccentricity consistent with the ring properties.}
\label{planetlimit}
\end{figure}

Interior to $\sim$ 30 $AU$ only a superjovian-mass planet can sculpt the ring by gap opening (Figure \ref{planetlimit}).  Planets with masses and semimajor axes comparable to the outer solar system planets could stir the disk to appear as a bright debris ring with an eccentricity of 0.1--0.2 (e.g. a Saturn with $e$ = 0.2).  In both cases, Super-Earths just interior to the ring edge could sculpt the ring.  
\subsection{Spectral Analysis}

We perform two spectral extractions.  First, we extract the raw spectrum at both disk ansae, identifying the `centroid positions from the collapsed image.  We then extract the spectrum of the best-fit disk model at these centroid positions.  In both cases, we measure the surface brightness: the mean brightness within the same 8-pixel diameter aperture used for flux calibration with errors calculated using method 2 described in Section 3.   For the model disk, we average the results from the two ansa and correct for signal loss due to a finite aperture by comparing the surface brightness of the convolved and unconvolved best-fit disk models.  

\begin{figure}
\centering
\includegraphics[scale=0.5,clip]{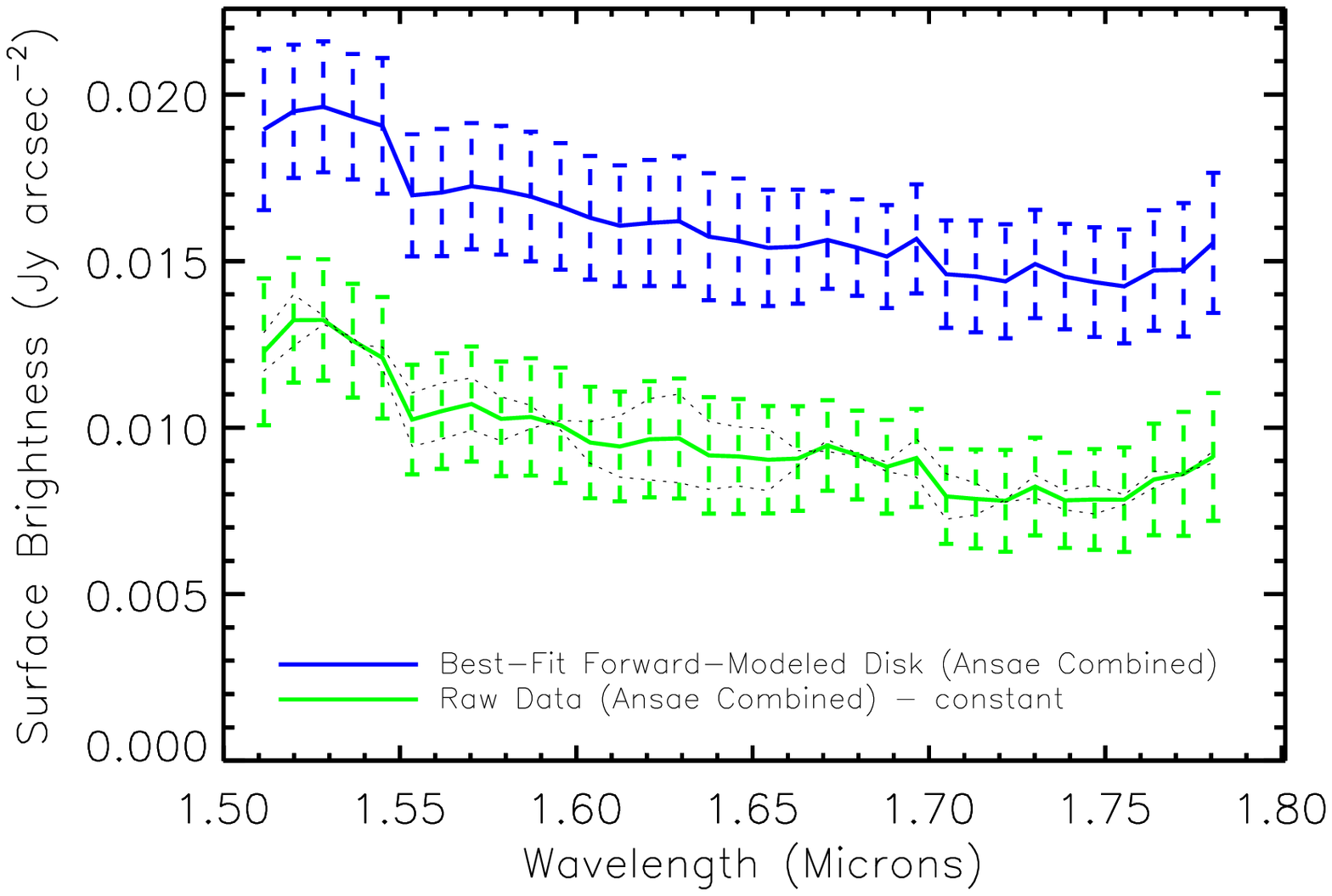}
\includegraphics[scale=0.5,clip]{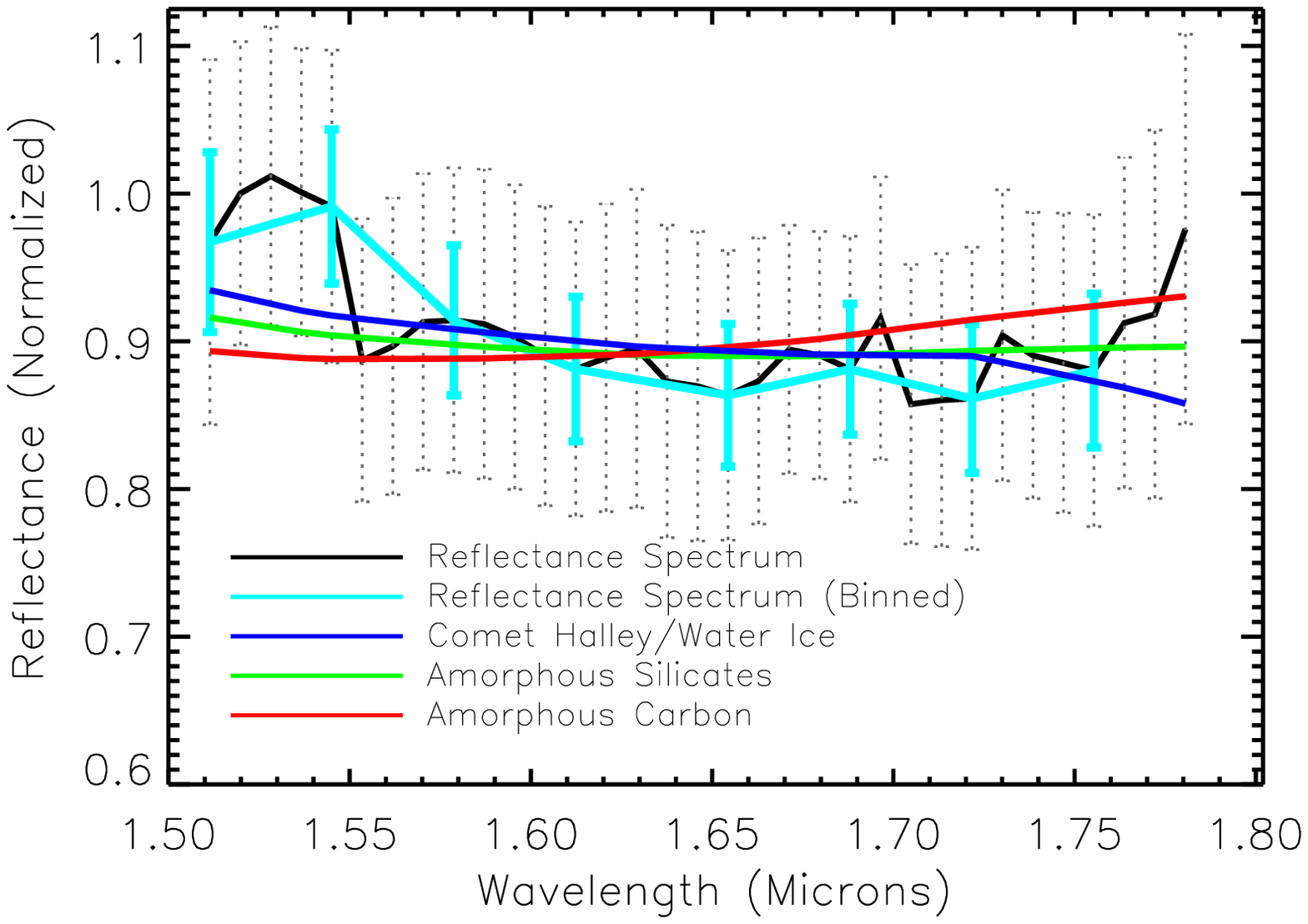}\\
\includegraphics[scale=0.6,clip]{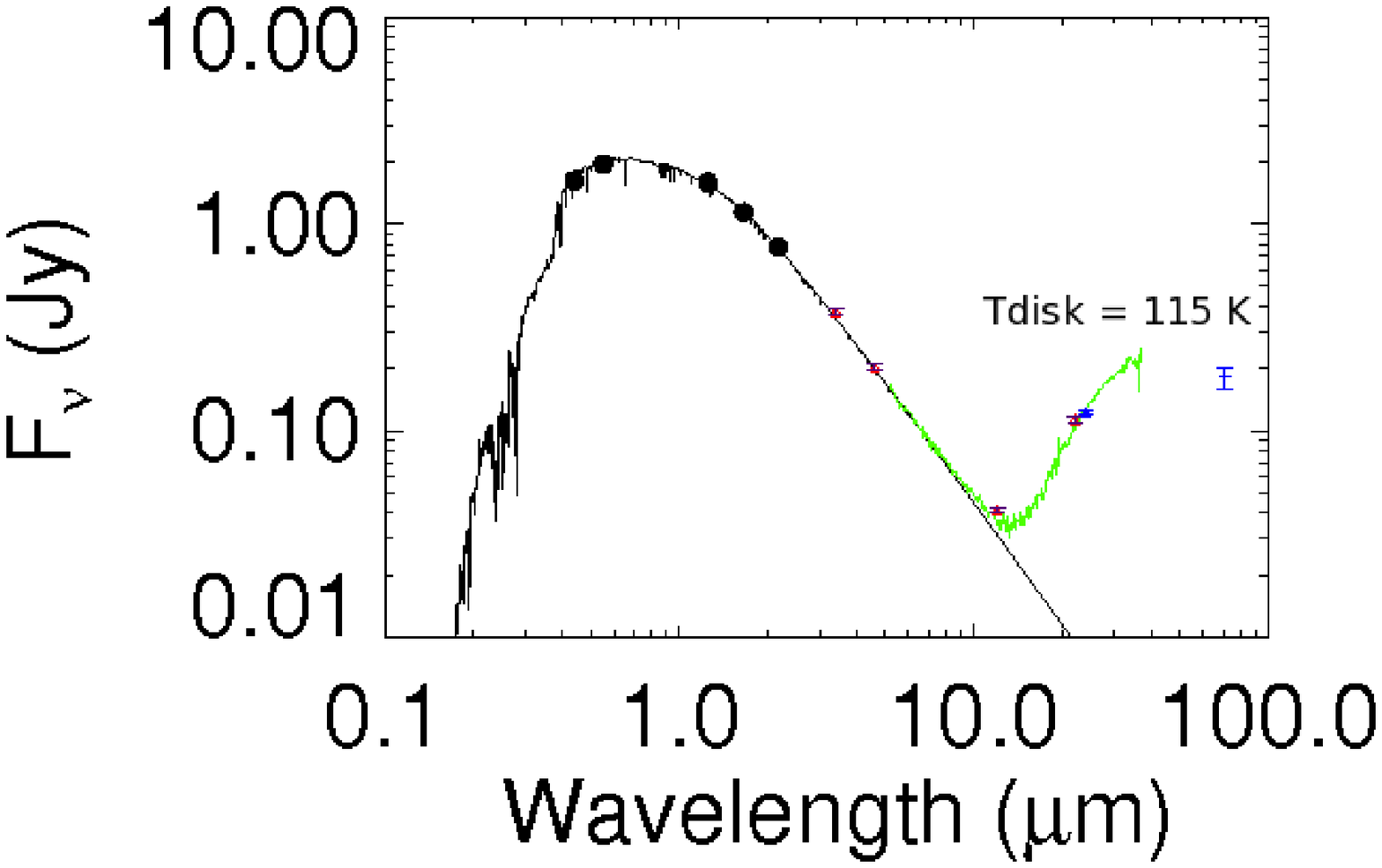}
\caption{(Top-Left) Surface brightness of the best-fit model disk spectrum (averaged between ansa).   Due to low throughput, we trim the first and last two channels.  
 The raw disk spectrum shows an almost identical shape but consistently $\sim$ 15\% lower signal: spectra extracted from individual ansae (dashed curves) agree within their error bars.  (Top-Right) Reflectance spectra of the best-fit model disk  and a binned (to the spectral resolution of GPI) version of the best-fit model compared to Mie theory predictions for water ice, amorphous silicates, and amorphous carbon.
 (bottom) HD 115600's optical/IR SED \citep{Chen2015} revealing 115 K $\pm$ 6 K (2-$\sigma$) cold dust and favoring an icy/ice-silicate composition (see main text).}
\label{spectra}
\end{figure}

Figure \ref{spectra}a shows our extracted best-fit scaled model disk spectrum (blue), averaged raw spectrum (green), and raw spectra extracted for each ansa (dotted lines).    The  spectra have a relatively flat to slightly blue slope across $H$ band.
Spectra extracted from the two ansae exhibit strong agreement.

The debris disk's scattered light spectral shape is sensitive to its dust's composition.  
 Instead of using advanced methods \citep[e.g.][]{Lisse2007}, we model the reflectance of dust using simple Mie theory-based single constituents found amongst dust and large bodies in the Kuiper belt-- carbonaceous dust, amorphous silicates, and water ice -- drawn from \citet{Lisse1998}, nominally using the best-fit Halley-like particle size distribution \citep{Krishna1988,Mazets1986} but exploring simple power laws with exponents between -3 and -4.   To calculate reflectance, we divide the best-fit model disk's spectrum by that of a Pickles F2V star.

The raw (black curve) and binned (cyan curve) disk reflectance spectra appear neutral/slightly blue at 1.5--1.75 $\mu$m (Figure \ref{spectra}b).   Reflectance spectra of water ice (blue), amorphous silicates (green) and amorphous carbon (red) show differences across the $H$ passband.   Water ice (and silicates) with a Halley-like size distribution provide the best match, reproducing the spectrum's neutral/blue slope.  Amorphous carbon has a $\chi^{2}$ value 2.5 times higher than water ice's and is marginally disfavored (at the 68\% confidence limit) for the binned spectrum, having a goodness-of-fit statistic of 0.78.  However, adopting a simple power law exponent for the particle size distribution between $-$3 and $-4$, a subset of models from each species  likewise fit.  
Thus, our spectrum's large error bars and narrow wavelength range currently preclude us from making definitive statements about the dust composition.  


Comparing the $H$-band spectrum with thermal IR data and properties of other debris rings may better reveal evidence for ice.   HD 115600's thermal disk emission (Figure \ref{spectra}c) peaks at 115 K, but blackbody-like emission should be located at 13 AU, not the observed 48 AU.  Real dust around stars later than A0 should be located at increasingly larger distances than predicted by blackbody emission, over 2.5 times further for an F2 star \citep[see Figure 10 in ][]{Booth2012} or beyond 33 AU in HD 115600's case.    Steeper size distributions and sub-blowout dust sizes (expected for a collisionally-active disk) may be sufficient to yield dust around HD 115600 at the right distances.
However,  organics (carbon) dominated dust yields dust locations an additional factor of  4--8 larger than predicted by blackbody emission; ice and ice/silicate mixtures yield negligible enhancement \citep[Figure 12 in][]{Booth2012}.  A water-ice (dominating scattered light) and organics (dominating thermal emission) mixture, much like that found on Kuiper belt object surfaces, may fit as well.
   
HR 4796A's fractional luminosity is over 2 times higher \citep[$\sim$ 4.8 $\times$ 10$^{-3}$;][]{Low2005, Chen2011}.  The $H$-band surface brightness contrast of HD 115600's disk ansae is $\sim$ 1.6 magnitudes brighter than HR 4796A's \citep{Rodigas2015}.  Moving the HR 4796A ring from 70 AU to HD 115600's disk location (48 AU) still results in scattered light emission about half as bright as HD 115600's disk.  Thus, HD 115600's disk is reflecting light more efficiently than HR 4796A's disk while having less thermal emission, a result explicable if HD 115600s disk is dominated by higher albedo species like water-ice.   Multi-wavelength photometry/spectroscopy is needed to more decisively assess the composition of HD 115600's disk.

\section{Discussion}
HD 115600's debris ring is the first object discovered using the new generation of extreme adaptive optics instruments \citep[SCExAO, GPI, and SPHERE;][]{Martinache2009,Macintosh2014,Beuzit2008} and the first debris ring with spatially-resolved integral field spectroscopy.   The ring is confined to a Kuiper belt-like stellocentric distance and its reflectance is consistent with that of major constituents of Kuiper belt bodies and their ejecta, including water ice.   As the disk orbits a 15 Myr-old star only $\sim$ 40--50\% more massive than the Sun located in the nearest OB association, it provides a promising reference point for understanding the early evolution and composition of the Kuiper belt.  

HD 115600's disk shows evidence for dynamical sculpting by a solar system-like giant planet.  With the exception of $\beta$ Pic b \citep{Lagrange2010}, fully-formed young planets directly imaged thus far have super-jovian masses (5--10 $M_{J}$) and orbit at wide separations (15--150 AU) \citep{Marois2008,Rameau2013,Kuzuhara2013,Currie2014a}.  Southern hemisphere  systems GPI and SPHERE could image even lower-mass planets located just interior to HD 115600's debris ring and superjovian planets near their inner working angles ($r$ $\sim$ 0\farcs{}1--0\farcs{}2 or $\sim$ 10--20 AU).  Combining new planet detections/upper limits with more precise estimates of the disk's properties will make HD 115600 an excellent laboratory for studying planet-disk interactions \citep[][]{Chiang2009,Rodigas2014}.  

\textbf{Acknowledgements} -- We thank the anonymous referee, Wladimir Lyra, Eric Mamajek, and Mengshu Xu for helpful comments; Jean-Charles Augereau for use of GRATER; Fredrik Rantakryo for executing these queue-mode observations; and the GPI Early Science Time Allocation Committee and Gemini Director Markus Kissler-Patig for supporting this program.  


{}

\begin{deluxetable}{lcllccccccc}
\setlength{\tabcolsep}{0pt}
\tablecolumns{4}
\tablecaption{Debris Disk Forward Modeling}
\tiny
\tablehead{{Parameter}&{Model Range}&{Best-Fit Model} & {Well-Fitting Models} }
\startdata
$r_{o}$ (AU)& 47 ... 48 & 48 & 47 ... 48\\
$e$  & 0 ... 0.3 & 0.2 & 0.1 ... 0.2\\
$\alpha_{in}$ & 7.5 ... 10 & 7.5 & 7.5 ... 10  \\
$\alpha_{out}$ &$-5$ ... $-7.5$ & $-7.5$ & $-5$ ... $-7.5$\\
$\Delta$y (AU) & 0 ... 1 & 1 & 0 ... 1\\
$g$ & 0 ... 0.15 & 0 & 0\\
$ksi_{o}$ (AU) & 1... 5 & 3 & 1 ... 5\\
\enddata
\tablecomments{We define the ``best fit" model as the one minimizing $\chi^{2}$ over $i$ convolved pixels where 
$\chi^{2}$ = $\sum\limits_{i}$ (model$_{i}$-image$_{i}$)$^{2}$/$\sigma_{i}^{2}$.  
}
\label{diskmodels}
\end{deluxetable}

\end{document}